# Growth, characterization, and magnetic properties of a Li(Mn,Ni)PO$_4$ single crystal

Kunpeng Wang[a,*], Andrey Maljuk[b], Christian G. F. Blum[b], Thomas Kolb[a], Carsten Jähne[a], Hans-Joachim Grafe[b], Lars Giebeler[b], Hans-Peter Meyer[c], Sabine Wurmehl[b], Rüdiger Klingeler[a,†]

[a]Kirchhoff Institute for Physics, University of Heidelberg, D-69120 Heidelberg, Germany

[b]Leibniz Institute for Solid State and Materials Research (IFW) Dresden, D-01171 Dresden, Germany

[c]Institut für Geowissenschaften, University of Heidelberg, D-69120 Heidelberg, Germany

**Abstract:**

Ni-doped LiMn$_{0.95}$Ni$_{0.05}$PO$_4$ single crystals have been grown for the first time by the travelling-solvent floating-zone method at low Argon pressure. The grown sample exhibits large single crystalline grains as revealed by means of polarization microscopy and X-ray Laue back scattering. The composition of the crystal was determined by Energy-dispersive X-ray (EDX) spectroscopy. LiMn$_{0.95}$Ni$_{0.05}$PO$_4$ orders in an orthorhombic olivine-like structure as expected and phase purity was confirmed by powder X-ray diffraction. An oriented cuboid with size of 2.4 × 2.5 × 2.7 mm$^3$ along *a*, *b*, and *c* crystalline directions, respectively, was used for anisotropic magnetic measurements.

[*] Kirchhoff Institute for Physics, INF 227, D-69120 Heidelberg, Germany; Email: kunpeng.wang@kip.uni-heidelberg.de
[†] Kirchhoff Institute for Physics, INF 227, D-69120 Heidelberg, Germany; Tel. +49 6221 54-9199 (fax: 9869); Email: klingeler@kip.uni-heidelberg.de

# 1. Introduction

While all end-member olivine phosphates LiMPO$_4$ (M = Mn, Fe, Co, or Ni) exhibit antiferromagnetic spin order and a large magnetoelectric effect in the magnetically ordered phase, the interest in these materials has been further boosted by the recent discovery of unusual ferrotoroidic domains in LiMPO$_4$ with M = Co or Ni [1,2]. In addition, olivine phosphates exhibit exceptionally high applicability for electrochemical energy storage in lithium-ion secondary batteries [3]. With regard to potential application in electrochemical energy storage, mixed transition metal ion compounds Li(M,M')PO$_4$ are particularly promising cathode materials since the redox voltage increases as from Fe to Mn, Ni, Co [4]. Indeed, it has been shown recently that LiMnPO$_4$ provides a competitive next generation cathode material which has been shown for carbon-coated nanostructured material with stable reversible capacity up to 145 mAh/g and a rather flat discharge voltage curve at 4.1 V [5,6]. Replacing Mn by Ni not only increases the cell voltage but also enhances the electrochemical performance in general by attenuating the distortion of the lattice while in turn improving the kinetic properties [7]. Therewith, Li(Mn,Ni)PO$_4$ may provide a next step towards high-energy cathode materials.

For detailed studies of the abovementioned phenomena, i.e. changes of the magnetic anisotropy and magnetoelectric coupling, or the evolution of the ferrotoroidic phase upon replacing of Mn by Ni in LiMnPO$_4$ on the one hand or of the anisotropic electric transport and Li diffusion on the other hand, the growth of large and high-quality single crystals is crucial. However, though the investigation of LiMPO$_4$ single crystals will shed light on the strategy to further optimize those functional materials, only few studies dealing with single crystals have been done so far and doping series with mixed transition metal ions are completely missing. Most of them applied polycrystalline powder or LiCl-flux grown crystals of LiFePO$_4$, LiMnPO$_4$ and LiNiPO$_4$ which have been available for a few years already [8,9,10,11]. For LiFePO$_4$, Chen et al. successfully applied the traveling-solvent floating-zone (TSFZ) technique with relatively fast growth rates between 2 and 4 mm/h and an Ar flow of 300 ml/min for minimizing the volatilization Li and to obtain a large stoichiometric crystal [12]. Noteworthy, the TSFZ- and flux-grown crystals provide qualitatively contradicting information on the nature of the transport properties, i.e. whether electronic/ionic conductivity and Li-diffusivity are essentially either one- or two-dimensional [10,12]. In addition, the effect of doping on the ionic conductivity has been studied on TSFZ-grown Al-doped and Si-doped single crystals [13,14]. LiCoPO$_4$ single crystals have been grown at Ar pressure of 7 bar [15] while LiMnPO$_4$ TSFZ-grown single crystals were produced not below 40 bar Ar pressure [16]. Here, we report for the first time on the growth of single crystalline LiMn$_{1-x}$Ni$_x$PO$_4$ with $x = 0.05$ at low Ar pressure and its initial characterization. The Ni-doped crystals will allow investigating the changes of magnetic anisotropy and magnetoelectric coupling which strongly differ in LiMnPO$_4$ and LiNiPO$_4$. The total rod size length amounts 50 mm with a diameter of 6.0 - 7.8 mm. Two cube-like single crystals of 10 – 15 mm$^3$ volume have been extracted from the rod for further studies (see Fig. 1). Phase purity and crystallinity is confirmed by powder X-ray diffraction as well as by polarized light imaging. Excellent crystal quality is confirmed by a highly anisotropic antiferromagnetic phase transition at $T_N = 32.5$ K and a very sharp spin-flop-like magnetic field-driven reorientation in the magnetically ordered phase. While the small level of Ni-doping already yields a significant reduction of B$_{SF}$ as compared to LiMnPO$_4$, the easy magnetic axis is still parallel to the crystallographic *a* direction. The material at hand therefore provides a first step to the spin reorientation towards the magnetically easy *c*-axis and the formation of the ferrotoroidic phase at high Ni-doping levels.

2. **Experimental Section**

## 2.1 Crystal Growth

Polycrystalline LiMn$_{0.95}$Ni$_{0.05}$PO$_4$ was prepared by a solid-state reaction using a stoichiometric mixture of Li$_2$CO$_3$ (Chempur 99+%), MnCO$_3$ (Aldrich 99.9+%), NiO (Aldrich 99.9+%) and NH$_4$H$_2$PO$_4$ (Chempur 99+%) under flowing argon, respectively. The chemical reaction can be expressed as:

$$\text{Li}_2\text{CO}_3 + 1.9\ \text{MnCO}_3 + 0.1\ \text{NiO} + 2\ \text{NH}_4\text{H}_2\text{PO}_4$$
$$\rightarrow 2\ \text{LiMn}_{0.95}\text{Ni}_{0.05}\text{PO}_4 + 2.9\ \text{CO}_2\uparrow + 3\ \text{H}_2\text{O}\uparrow + 2\ \text{NH}_3\uparrow$$

The mixture of the precursor materials was ground carefully to ensure homogeneity, and then the powder was sintered for 20 h at 800 °C. Subsequently, the product Ni-doped was pressed at 2 kbar (CIP, Engineered Pressure Systems; 2000 bar) and sintered again at 800 °C for 10 h. The LiMn$_{0.95}$Ni$_{0.05}$PO$_4$ crystal was grown in a floating zone facility with IR optical heating (Crystal System Incorporation, Japan). The four 300 W air-cooled halogen lamps were employed. A quartz tube with 2 mm wall thickness was used as the growth chamber. The applied Ar pressure in the growth chamber was 2 bar. The feed rod was rotated clockwise at a rate of 15 rpm, and the seed anticlockwise 15 rpm. The growth rate was 2.5 mm/h.

## 2.2 Crystal Characterization

The elemental concentration of LiMn$_{0.95}$Ni$_{0.05}$PO$_4$ was analyzed on a scanning electron microscope (SEM) LEO 440 with Oxford Inca X-Max 80 detector. Accelerating voltage was 20 kV, working distance appr. 25 mm, counting time 100 seconds (livetime) at about 10.000 cps.

Powder X-ray diffraction data were collected on a STOE Stadi P diffractometer equipped with a 6°-linear position sensitive detector and a curved Ge(111)-monochromator of the Johann-type. Samples were measured with Cu K$_{\alpha 1}$-radiation in the range of 5° ≤ 2$\theta$ ≤ 90° in transmission geometry (flat sample) with a step size of Δ2$\theta$ = 0.01°. The powder sample was prepared as a thin layer on a polyacetate film with a collodium/amyl acetate glue. For the Rietveld refinement the software package WinPlotR including Fullprof was used [17]. As structure model, the structure data of undoped LiMnPO$_4$, space group *Pnma* [18], was implemented into the Fullprof program routine according to the standardized data set as provided by the ICSD [19]. The microstructure and crystal perfection of several samples were investigated by optical microscopy in a polarization microscope Axiovert 25 equipped with adigital camera (both Carl Zeiss). The orientation of the single crystal was determined by the X-ray Laue back-scattering method, which was performed at a Seifert X-ray diffraction system with Mo K$_\alpha$ radiation. The power is 50 kW with a current of 40 mA. The exposure time on the film was 30 min.

The temperature dependence of the static magnetic susceptibility $\chi$ = M/B was measured both after cooling in zero magnetic field (ZFC) and in the actual probing field (FC), respectively. The data were collected at temperature from 2 to 380 K in an applied external magnetic field of B = 0.1 T applied along the three crystalline axes of an oriented cuboid by means of a Quantum Design MPMS XL5 SQUID magnetometer. The field dependence of the magnetization was studied at T = 1.8 K in magnetic fields up to 5 T.

## 3. Results and Discussion

Figure 1 shows the as-grown rod of LiMn$_{0.95}$Ni$_{0.05}$PO$_4$ with a 6 – 7.8 mm diameter and 50 mm length. From Figure 1 we can see that the material displays orange transparent color, which is the same as in undoped LiMnPO$_4$ [16]. The phase purity of the grown crystal was checked by powder X-ray diffraction (XRD) on a powderized piece of the single crystal (see Fig. 1). The resulting XRD pattern is shown in Fig. 4. All Bragg reflections are indexed by the structure of LiMnPO$_4$, space group *Pnma* [18]. Additional phases are excluded due to the absence of any additional reflection within the detection limit of the XRD method. The lattice parameters are determined to *a* = 10.4247(13) Å, *b* = 6.0906(8) Å, and *c* = 4.7358(6) Å. The unit cell volume of 300.69(11) Å$^3$ at room temperature is slightly lower than 301.12(3) Å$^3$ of polycrystalline LiMnPO$_4$ [18]. The reduced unit cell volume of the reported phase may point to some replacement of Mn$^{2+}$ by Ni$^{2+}$. When associating the Ni-doping with negative chemical pressure [20], our results imply $\Delta V/V \approx$ -1·10$^{-3}$/% Ni. There is no clear hint on a substitution of larger amounts of Mn by Ni owing the small total amount of nickel. Considering in addition that pure LiNiPO$_4$ ($V$ = 274.49(1) Å$^3$) [18] and pure LiMnPO$_4$ form isostructural compounds, no large influence of Ni on the structure is expected. Some Bragg reflections are underestimated by the refined structure model corresponding to remaining single crystal contributions to the powder pattern. These contributions can result from preferred orientations respectively the shape of the grains due to the crystal growth process.

For the cross-sectional polarized light images, a 8 mm thick sample was cross cut from the initial part of the LiMn$_{0.95}$Ni$_{0.05}$PO$_4$ crystal rod. The surface of the cylindrical sample toward to the initial part was labeled as Q5, and the other side of the sample was labeled as Q6. The optical images of the Q5 and Q6 samples in polarized light are shown in Figure 2(a) and 2(b), respectively. The transmittance of the polarized light is in most parts homogeneous indicating that the as-grown crystal boule is consisting of mainly one large grain throughout the entire cross-section. The few areas with high transmission on both surfaces were induced by local stress, which presumably can be eliminated by thermal annealing. Besides the local stress, we also observe a few mechanical scratches and grain boundaries. No significant difference can be observed on both sides of the sample (Q5 and Q6), indicating that the growth conditions of the crystal such as the applied power, growth rate and speed of rotation are relatively stable. The as-grown crystal quality is apparently good as the common macro-defects such as cracks, precipitations, inclusions, striations and bubbles were not observed. To further investigate the defects in the grown crystal, we cut the samples into slices with thickness lower than 2 mm (cf. Fig. 1). The polarized microscopic image is shown in Figure 2(c). Regular interference fringe can be obviously observed, implying that the optical homogeneity of the crystal is good.

The X-ray Laue back-scattering image, which was obtained in the position represented by the white circle in Fig. 2a, of the 8 mm thick cylindrical sample shown in Fig. 2d is discussed exemplarily below. Similar Laue experiments were done at different spots of all surfaces of the cuboid used for the magnetization experiments (see Fig. 5), giving the same results, which further confirms large single crystalline grains. The Laue data obtained at the white spot region imply that the growth direction does not coincide to one of the crystal main axes. The Laue spots can be clearly observed. Seven diffraction peaks with the highest intensities were selected to simulate the geometric distribution of the Laue spots based on the orthorhombic lattice unit of LiMnPO$_4$. The simulated Laue pattern shown in Figure 2d` agrees well to the experimental one, confirming the orthorhombic symmetry. All the diffraction spots were then indexed successfully confirming that the structure of the sample is basically the same than in undoped LiMnPO$_4$. The Laue spot located in the center of the film is indexed as (2 1 $\bar{1}$), which is corresponding to the lattice plane parallel to the film. As the sample was cross cut from the LiMn$_{0.95}$Ni$_{0.05}$PO$_4$ crystal rod, we can deduce that the crystal was grown along the $\langle 2\ 1\ \bar{1} \rangle$ crystalline direction. The Laue patterns were used to cut the sample along the main crystal directions, as e.g. exemplified in Figures 2e and 2e` showing orientation perpendicular to the (100) plane.

Chemical analysis was performed by means of energy-dispersive X-ray analysis. In order to minimize the measurement uncertainty, three different areas (50 x 50 μm) have been measured for deriving the chemical composition. For the calibration pure LiMnPO$_4$ was used as internal standard. The Li content generally cannot be analyzed by a standard EDX measurement. Oxygen was not determined from the spectra but assumed to be stoichiometric with the cations, Mn$^{2+}$, Ni$^{2+}$, and P$^{5+}$. Accordingly, when normalizing to 100 %, Li$_2$O was neglected. The calculation of the stoiciometric formula is based on 3.5 oxygen atoms per formula unit. The typical EDX spectra exhibit Mn, Ni, P, and O peaks, in addition to the C peak stemming from the coating layer. The results of the analysis are shown in Table 1. For Mn, Ni, and P the atomic ratios nearly match the nominal stoichiometry, i.e. 0.952 Mn, 0.058 Ni, and 0.996 P. The statistical uncertainty as determined by the root mean square deviations from the average values is negligible, i.e. ±0.001 for Mn, Ni, and P. The average Mn:Ni ratio amount to 94.23:5.77 which is in a good agreement with the initial weight ratio and there are only small deviations at the different spots. This indicates that all the initially used Ni was incorporated in the main phase of Li(Mn,Ni)PO$_4$ and a rather homogenous Ni distribution.

Measurements of the static magnetic susceptibility $\chi$ = M/B vs. temperature are displayed in Fig. 5. For these measurements, a cuboid with size of 2.4 × 2.5 × 2.7 mm$^3$ along the crystallographic *a*, *b*, and *c* direction was oriented and cut from the above mentioned cylindrical as-grown sample (compare Fig. 1). Note, that the inset of Fig. 5 also shows the magnetic susceptibility of a powderized piece of the single crystal for comparison. The data imply a paramagnetic Curie-Weiss-like behavior at high temperatures and long range antiferromagnetic order below $T_N$ = 32.5 K. At higher temperatures, the data can be described by applying the Curie-Weiss-law $\chi(T) = \chi_0 + C/(T+\Theta)$, with the Curie-constant C, a temperature-independent term $\chi_0$ = -3·10$^{-4}$ emu/(G$^2$mol) and the Curie-Weiss temperature $\Theta \approx$ -70 K. Fitting the data yields an effective magnetic moment of $p$ = 5.99 $\mu_B$. Analyzing the magnetic specific heat which is proportional to $\partial(\chi T)/\partial T$ (see upper inset of Fig. 6) shows that the onset of long-range magnetic order is associated with a pronounced and very clear lambda-like anomaly. In addition, there is a pronounced anisotropy of the magnetic susceptibility in the magnetically ordered phase which again signals the high quality of the crystal. This is also visible in the magnetic field dependence of the magnetization shown in Fig. 6. Upon application of B parallel to the easy magnetic *a*-axis, the data exhibit a clear and sharp spin-flop transition at $B_{SF}$ = 2.8 T. At this field, a spin-reorientation occurs so that the magnetic moments are mainly aligned perpendicular to the external field. Please note that this value is clearly reduced as compared to undoped LiMnPO$_4$ where $B_{SF}$ = 4 T [21]. For B||a > $B_{SF}$, the magnetization curve is linear and extrapolates to zero, as expected for a spin-flop transition. Noteworthy, the M(B||b) curve is a straight line in the accessible field range

which again is expected for an Neel-type antiferromagnetic order. However, any non-ordered impurity would cause a Brillouin-like field dependence, i.e. a right curvature at low-fields and fast saturation. The M(B||a) data at hand do not exhibit a clear non-linear contribution. Analyzing the data allows to quantitatively estimate the upper impurity limit to be about $\approx 2\cdot 10^{-4}$ $\mu_B$/f.u. only.

## 4. Conclusions

A LiMn$_{0.95}$Ni$_{0.05}$PO$_4$ single crystal has been grown by the floating zone technique. The grown crystal was carefully characterized by polarized microscopic images, X-ray Laue back scattering technique, X-ray powder diffractions, and the EDX method. An oriented cuboid with size of 2.4 × 2.5 × 2.7 mm$^3$ along *a*, *b*, and *c* crystalline directions was used for studies of the anisotropic magnetization.

**Acknowledgements**

The authors thank I. Glass, R. Müller, C. Malbrich, and K. Leger for experimental assistance. Support by the Bundesministerium für Bildung und Forschung (BMBF) within project 03SF0397 and by Deutsche Forschungsgemeinschaft DFG via KL1824/5-1, WU595/3-1, and GR3330/2-1 is gratefully acknowledged. KW acknowledges support by the European Commission through FP7 Marie Curie grant PIIF-GA-2012-331476 LiCrystG.

**Figure captions**

1. The as-grown LiMn$_{0.95}$Ni$_{0.05}$PO$_4$ single crystal. Q5 and Q6 mark cuts mentioned in the text, the circle indicates the position where the material for the powder XRD has been taken from, and two small thin crystal pieces were used for additional optical imaging (see the text). The oriented cuboid was extracted as sketched.

2. Optical images in polarized light of both the (a) Q5, and (b) Q6 cross sections cut from the initial part of the LiMn$_{0.95}$Ni$_{0.05}$PO$_4$ crystal, as well as the polarization image of (c) a 2 mm thick plate; X-ray Laue back scattering images of the (d) experimental, and (d`) simulated patterns with orientation, as well as the (e) experimental, and (e`) simulated patterns with axis. The Laue pattern was determined on the position labeled by the white circle in part (a) of this figure.

3. Energy dispersive X-ray spectrum of the LiMn$_{0.95}$Ni$_{0.05}$PO$_4$ crystal.

4. X-ray powder diffraction pattern of a powderized LiMn$_{0.95}$Ni$_{0.05}$PO$_4$ single crystal. As reliability factors R$_p$ = 1.70, R$_{wp}$ = 2.17, R$_e$ = 2.03 and $\chi^2$ = 1.14 are calculated.

5. Static magnetic susceptibility $\chi$ = M/B obtained on the cuboid shown for external magnetic field oriented along the three main crystal axes. The line represents a Curie-Weiss approximation to the average susceptibility $(\chi_a+\chi_b+\chi_c)/3$, the inset highlights the low temperature region and displays the susceptibility obtained on a powder sample.

6. Field dependence of the magnetization at T = 2 K for B∥a and B∥b. The upper inset shows the magnetic specific heat as calculated from the static susceptibility $\chi_{B∥b}$(T), the lower inset presents the derivative $\partial M/\partial B$ of the data shown in the main plot.

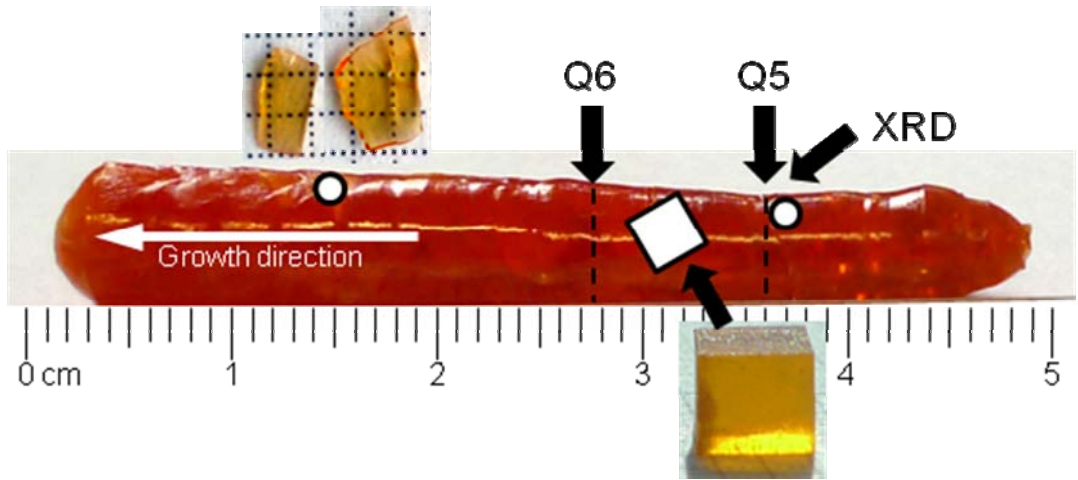

Figure 1

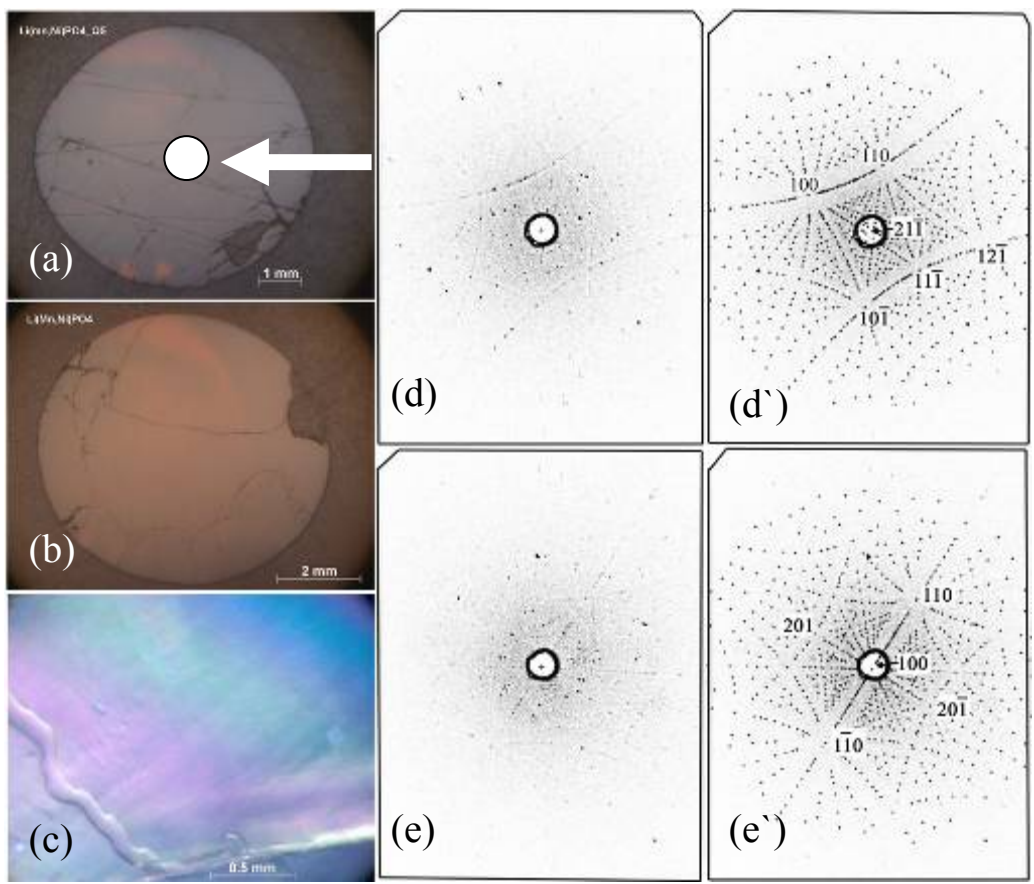

Figure 2

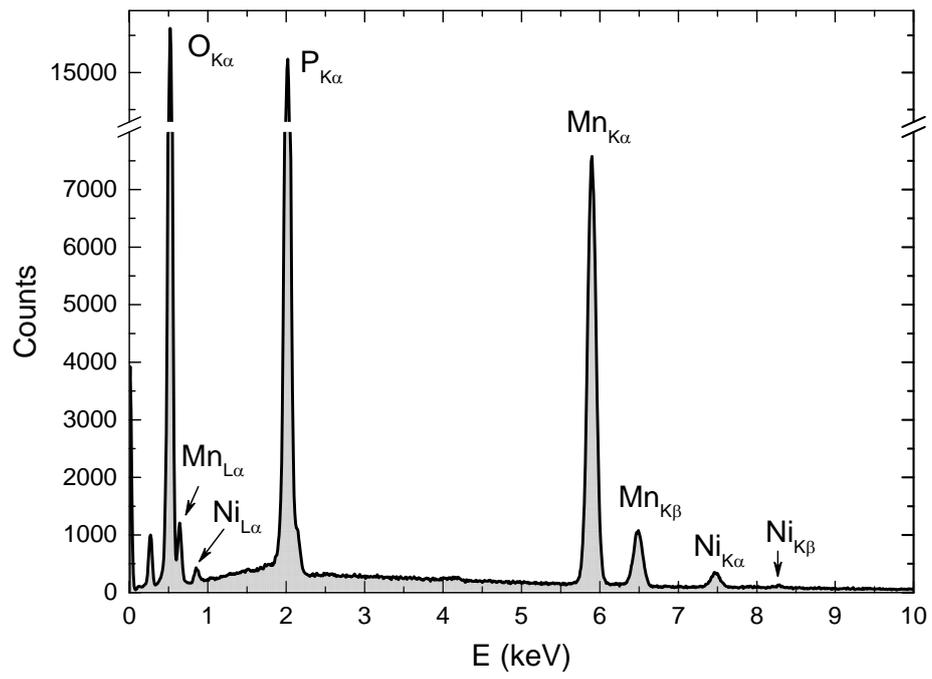

Figure 3

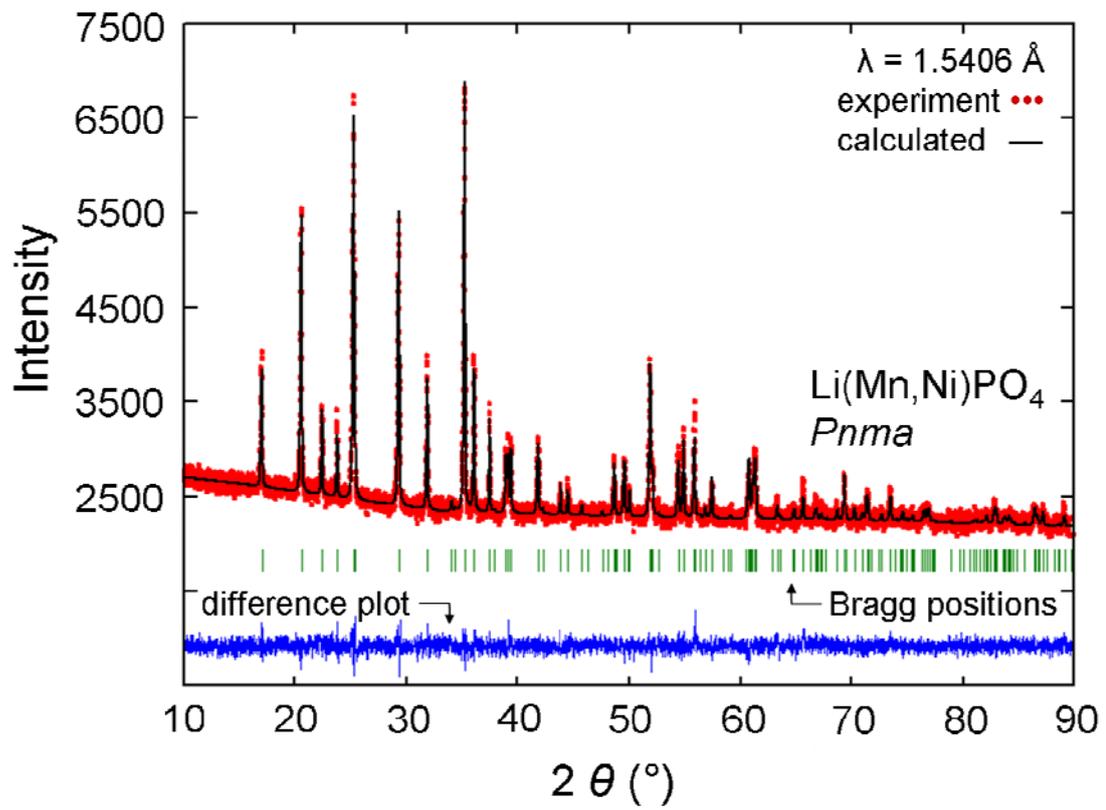

Figure 4

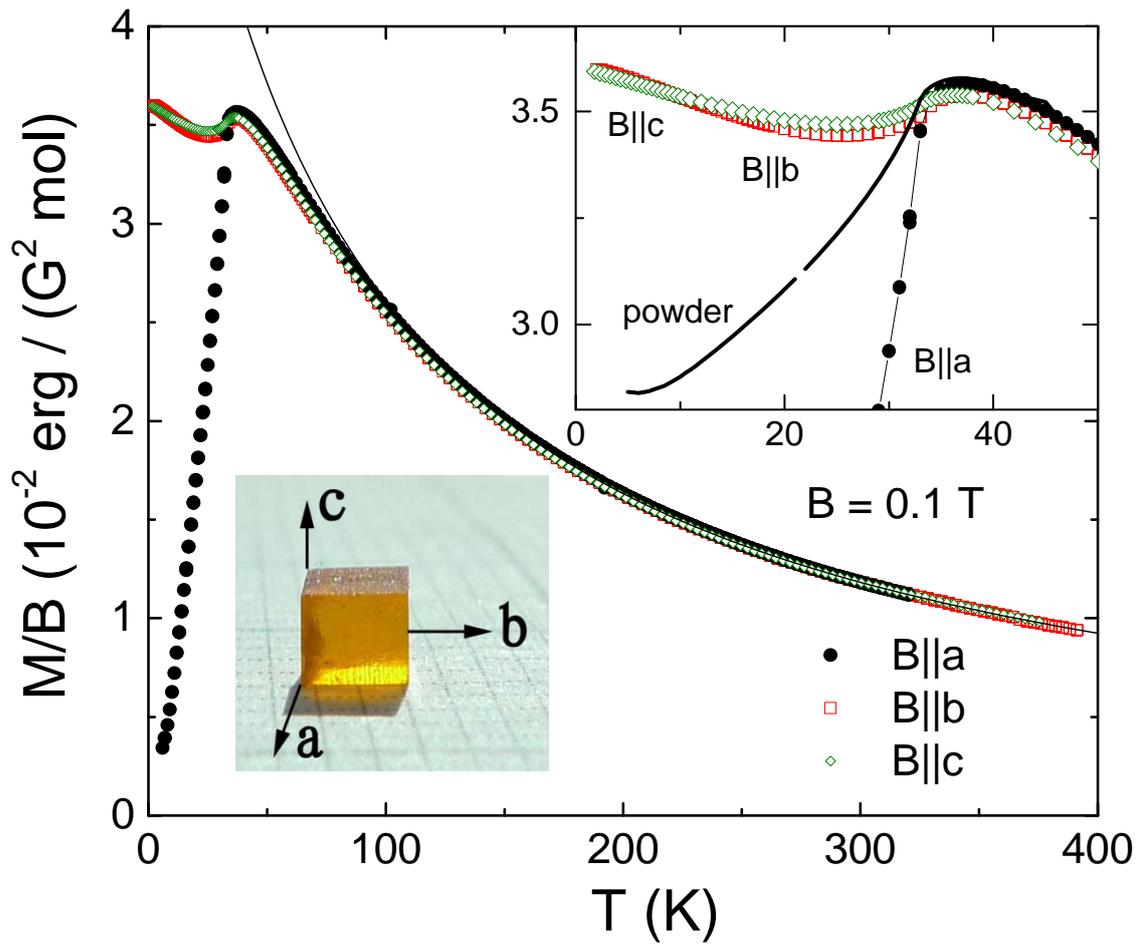

Figure 5

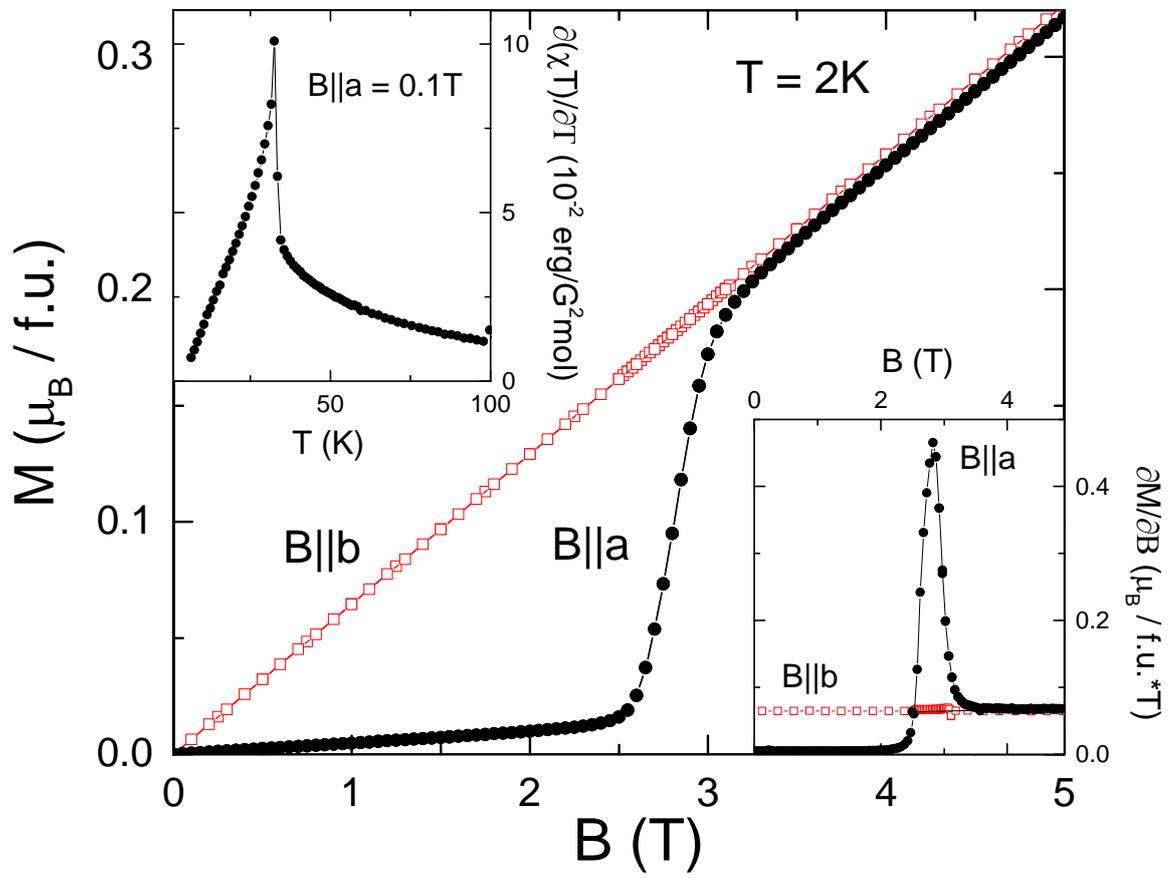

Figure 6

**Tables:**

Table 1: The chemical composition as determined by EDX analysis on three areas at the (001) surface of the LiMn$_{0.95}$Ni$_{0.05}$PO$_4$ crystal. The table shows the experimental data with respect to a pure LiMnPO$_4$ standard in wt%, the average Ø, the root mean square RMS, and the calculated numbers of Mn, Ni, and P cations based on 3.5 oxygens per formula unit (see the text).

|  | Wt% | | | | |  | Number of cations |
|---|---|---|---|---|---|---|---|
|  | area 1 | area 2 | area 3 | Ø | RMS |  | Ø |
| MnO | 47.46 | 47.29 | 47.36 | 47.37 | 0.07 | Mn | 0.952 |
| NiO | 3.13 | 3.03 | 3.01 | 3.06 | 0.05 | Ni | 0.058 |
| P$_2$O$_5$ | 49.41 | 49.69 | 49.64 | 49.58 | 0.12 | P | 0.996 |
| Sum | 100.00 | 100.01 | 100.01 | 100.01 |  | Sum | 2.006 |

# References

1. B.B. van Aken, J. P. Rivera, H. Schmid, M. Fiebig, Nature 449, 702 (2007)
2. R. Toft-Petersen et al., Phys. Rev. B 85, 224415 (2012)
3. J.-M. Tarascon, M. Armand, Nature 414, 359 (2001); S.-Y. Chung, J. T. Bloking, Y.-M. Chiang, Nature Mater. 1, 123 (2002); S. P. Herle, B. Ellis, N. Coombs, L. F. Nazar, Nature Mater. 3, 147 (2004)
4. A. Vadivel Murugan,, T. Muraliganth, P. J. Ferreira, A. Manthiram, Inorg. Chem. 48, 946 (2009)
5. D. Wang et al., J. Power Sources 189, 624 (2009)
6. V. Aravindan, J. Gnanaraj, Y.-S. Lee, S. Madhavi, J. Mater. Chem. A 1, 3518 (2013)
7. M. Minakshi, S. Kandhasamy, Curr. Op. Solid State Mater. Sci. 16, 163 (2012)
8. V. I. Formin, V. P. Gnezdilov, V. S. Kurnosov, A. V. Peschanskii, A. V. Yeremenko, H. Schmid, J.-P. Rivera, S. Gentil, Low Temp. Phys. 28, 203 (2002)
9. P. R. Elliston, J. G. Creer, G. J. Troup, J. Phys. Chem. Solids 30, 1335 (1969)
10. J. Li et al. Phys. Rev. B 73 (2006) 024410
11. J. Li, T. B. S. Jensen, N. H. Andersen, J. L. Zarestky, R. W. McCallum, J.-H. Chung, J. W. Lynn, D. Vaknin, Phys. Rev. B 79, 174435 (2009)
12. D. P. Chen, A. Maljuk, C. T. Lin, J. Cryst. Growth 284, 86 (2005). R. Amin, J. Maier, P. Balaya, D. P. Chen, C. T. Lin. Solid State Ionics 179, 1683 (2008); R. Amin, P. Balaya and J. Maier, Electrochem. Solid-State Lett. 10, A13 (2007); R. Amin and J. Maier, Solid State Ionics 178, 1831 (2008)
13. R. Amin, C. T. Lin, J. Peng, K. Weichert, T. Acartürk, U. Starke, J. Maier, Adv. Funct. Mat. 19, 1697 (2009)
14. R. Amin, C. T. Lin, J. Maier, Phys. Chem. Chem. Phys. 20, 3519 (2008); ibid. 3524 (2008)
15. R. Saint-Martin, S. Franger, J. Cryst. Growth 310, 861 (2008)
16. N. Wizent, G. Behr, F. Lipps, I. Hellmann, R. Klingeler, V. Kataev, W. Löser, N. Sato, B. Büchner, J. Cryst. Growth 311, 1273 (2009)
17. T. Roisnel, J. Rodriguez-Carvajal, Mat. Sci. Forum 378-381, 118 (2001)
18. O. García-Moreno, M. Alvarez-Vega, F. García-Alvarado, J. García-Jaca, J. M. Gallardo-Amores, M. L. Sanjuán, U. Amador, Chem. Mater. 13, 1570 (2001)
19. R. Allmann, R. Hinek, Acta Cryst. A 63, 412 (2007)
20. R. D. Shannon, Acta Cryst. A 32, 751 (1976)
21. J. H. Ranicar, P. R. Elliston, Phys. Lett. A 25, 720 (1967)